\documentclass[12pt]{iopart}
\usepackage{graphicx}% Include figure files

\begin{document}

\title{A model of magnetic order in hexagonal HoMnO$_3$}

\author{S G  Condran and M L Plumer}

\address{Department of Physics and Physical Oceanography, Memorial University, St. John's, Newfoundland, Canada, A1B 3X7}

\ead{plumer@mun.ca}

\begin{abstract}
Symmetry arguments are used to develop a spin Hamiltonian for the description of the complex magnetic ordering in HoMnO$_3$. Using a novel application of the Landau Lifshitz Gilbert dynamic torque equations to this model of the frustrated Mn ions on an $AB$ stacked triangular antiferromagnetic, it is shown that the four principal spin configurations observed in this compound are stabilized.   Ho-Mn coupling is found to be a consequence of an unusual trigonal anisotropy term which is responsible for simultaneous Mn spin reorientation and onset of Ho magnetic order.  Based on these microscopic considerations, a mean-field Landau-type free energy is derived which reproduces the succession of observed temperature driven magnetic phase transitions at zero field, including re-entrant behavior. In addition, our analysis suggests that the basal-plane magnetic order should be slighty incommensurate with the lattice.
\end{abstract}

%\keywords{}

\maketitle
%\section{Introduction}
%\section{Introduction}

Hexagonal HoMnO$_3$ is one of the most studied of the rare-earth manganites and has a magnetic field - temperature phase diagram that exhibits a multitude of complex spin structures \cite{munoz01,fiebig02,vajk05,lorenz05,yen07,camirand09}.  Key features of the magnetic phases in this stacked triangular antiferromagnet include a series of reorientation transitions (with re-entrant behavior) at about $5 ~ K$ and $40 ~ K$ involving $AB$-stacked triangular layers of Mn spins which form the familiar basal-plane 120$^0$ spin structure below the N$\acute{e}$el temperature $T_N = 76 ~ K$.  Low  temperature $c$-axis spin alignment of $AA$-stacked Ho ions is also observed which is concomitant with Mn spin reorientation.  The nature of the coupling between Mn and Ho ordering has remained puzzling for close to two decades.  

In this Communication, a formulation of the spin Hamiltonian appropriate for a description of the magnetic phases of HoMnO$_3$ is developed on the basis of symmetry arguments.  The principal Mn spin configurations observed in the magnetic phase diagram are found to depend on the signs of the coefficients of in-plane sixth-order anisotropy and inter-plane exchange coupling.  These states are determined through a numerical solution of the dynamic Landau-Lifshitz-Gilbert (LLG) equation which includes finite temperature effects.  Theoretical arguments proposed over twenty years ago that even weak inter-layer coupling of $AB$ stacked triangular antiferromagnetic layers leads to an incommensurate distortion of the usual 120$^0$ period-3 spin structure \cite{saka84} are confirmed by these simulations.  The crucial role of Ho-ion ordering in driving the transitions between the various Mn-ion ordered states is also explicitly demonstrated \cite{nandi08}. Since strong planar anisotropy in the case of Mn spins  (${\bf S} \perp {\bf c}$) and axial anisotropy in the case of Ho spins (${\bf S}_0 \parallel {\bf c} \parallel {\bf {\hat z}}$) leads to ${\bf S} \cdot {\bf S}_0 =0 $, the usual exchange coupling mechanism should be absent in equilibrium (but can contribute to magnetic excitations \cite{talbayev08}).  However, the hexagonal crystal symmetry of this compound contains a 6-fold screw axis which allows for a fourth-order trigonal anisotropy term of the form \cite{simpson95}
\begin{eqnarray}
\mathcal{H}_{\tilde{K}} = \tilde{K} \sum_{\langle ij \rangle} S_{0i}^z S_j^y[3(S_j^x)^2 - (S_j^y)^2]
\end{eqnarray}
where the sum is over near neighbors.  Such a term can occur provided that either Ho or Mn spin configurations are antiferromagnetic between $AB$ layers, but not both.  The consequences of this interaction are examined within the framework of a simple Landau-type free energy derived within a mean-field approximation.  This demonstrates that the trigonal coupling term causes the series of Mn-spin reorientation transitions observed in zero applied magnetic field.   We note that a phenomenological Landau-type free energy based on group theoretic arguments was previously developed to explore the magnetic phase diagrams of hexagonal RMnO$_3$ compounds, which is complementary to the present work \cite{munawar06}.

Our investigation may be compared with other proposals regarding microscopic mechanisms of coupling between rare-earth and Mn ions in the hexagonal manganites.  A recent discussion of Mn-Yb coupling in the sister compound YbMnO$_3$ (which does not exhibit zero-field Mn-reorientation transitions) suggests that it could be due to dipolar or Dzyaloshinksii-Moriya (DM) interactions \cite{fabreges08}.  Dipolar effects were estimated to be an order of magnitude too small and either effect would require that the rare-earth ions order at the same temperature as the Mn ions (since both types of interactions involve a linear coupling between the two types of spins), which is not the case in HoMnO$_3$.

An intriguing  alternative explanation has been put forth which suggests that the slight distortion of the perfect triangular symmetry can give rise to asymmetric inter-layer exchange interactions \cite{munawar06,fabreges09}. A change in sign of this exchange coupling occurs in response to the temperature-driven lattice distortion which in turn drives the Mn-spin reorientations.  The significance of such distortion-driven interlayer-coupling may be considered secondary in view of the observation made here that ordinary inter-layer exchange should indeed be present (since the in-plane Mn modulation should not be exactly period-3).  In addition, it is well established that there are simultaneous changes in the Ho magnetic order at the Mn-spin reorientation transitions which must be accounted for in a complete microscopic model. We do show below, however, that the distortion of the perfect triangular lattice resulting in anisotropic in-plane exchange interactions serves to enhance the incommensurability of the Mn spin modulation (in addition to the ordinary exchange coupling between $AB$ stacked layers mentioned above) \cite{zhitomirsky96}.   

Terms which appear in the spin Hamiltonian (or Landau-type free energy) of a magnetic system can be constructed with the requirement of invariance with respect to the generators of the crystal space group in the paramagnetic regime.  In the case of hexagonal RMnO$_3$ compounds, the crystal symmetry group is non-symorphic $P6_3cm$ with generators given by a screw rotation $\{C_6^+|00\frac{1}{2}\}$ and glide plane $\{\sigma_v|00\frac{1}{2}\}$  \cite{plumer07}.  Terms involving lattice vectors (such as dipole contributions) are omitted here for simplicity. In the present case, allowed anisotropic terms at second order are the axial/planar type, $(S^z)^2$, as well as the DM interaction.   The lowest order term which gives rise to a dependence of the orientation of planar spins relative to the crystal axes occurs at sixth order. 

In the remainder of this Communication results are presented on LLG simulations involving only Mn spins and an analysis of a mean-field Landau-type free energy with Ho-Mn trigonal coupling.  We conclude with a discussion of incommensurabiltiy arising from distortions of perfect triangle.

With the goal exploring a simplified spin Hamiltonian involving only the Mn ions we consider the following contributions
\begin{eqnarray}
\mathcal{H}_{Mn} &=&  \sum_{\langle ij \rangle} J_{ij} {\bf S}_i \cdot {\bf S}_j - D\sum_j (S_j^z)^2 \nonumber \\
&+& E\sum_j [(S_j^x + iS_j^y)^6 + (S_j^x - iS_j^y)^6].
\end{eqnarray} 
The first term includes only isotropic in-plane near-neighbor exchange interaction of the perfect triangular lattice, $J \equiv 1$ as well as near-neighbor exchange between $A$ and $B$ triangular layers, $J'$. Interlayer exchange is assumed here to be weak (as in YMnO$_3$ \cite{sato03}).  As noted previously \cite{saka84}, this contribution is zero in the case of the 120$^0$ spin structure since the coupling between a spin on layer $A$ and its three neighbors on layer $B$ appears as $J'{\bf S}^A \cdot ({\bf S}^B_1 +  {\bf S}^B_2 +{\bf S}^B_3)$.  Strong planar anisotropy is assumed to arise from a large single-ion term with $D < 0$.  The sign of the in-plane anisotropy coefficient $E$ determines the orientation of ${\bf S}$ relative to the crystal ${\bf a}$-axis in a temperature regime below $T_N$. 
 
This model Hamiltonian was used to demonstrate the four principal Mn spin configurations identified in RMnO$_3$ compounds.  These are conveniently illustrated in Fig. 7. of Ref.~\cite{fabreges08} and are labeled $\Gamma_{1 \rightarrow 4}$ and can be characterized by two features: (1) inter-plane ferromagnetic or antiferromagnetic spin configurations, and (2) intra-plane spin orientations being parallel or perpendicular to a basal plane ${\bf a}$-axis.

Spin configurations were determined numerically as a solution to the dynamic LLG equations, which may be expressed as the the familiar torque equation, involving the gyromagnetic ratio $\gamma$, plus a damping term involving the parameter $\alpha$, as 
\begin{eqnarray}
\frac{d{\bf S}}{dt} = -\frac{\gamma}{1+\gamma} {\bf S} \times {\bf H}_{eff} - \frac{\alpha \gamma}{1+\alpha^2} {\bf S} \times ({\bf S} \times {\bf H}_{eff}).
\end{eqnarray}
The LLG equations have proven very useful in the determination of subtle differences in equilibrium magnetic structures which involve long-range magnetostatic effects in thin films but have occasionally also been applied to spin systems with only short-range exchange interactions  \cite{feng01,szunyogh09}. They are well suited for the study of highly frustrated magnetic systems and may be viewed as an alternative to Monte Carlo simulations.  The effective field is given by ${\bf H}_{eff} = - \delta \mathcal{H} / \delta {\bf S} + {\bf H}_{th}$ where ${\bf H}_{th}$ is the stochastic field describing thermal fluctuations within the framework of Langevin dynamics.
Spin vectors were located on interpenetrating $12 \times 12 \times 12$ $AB$ stacked triangular lattices. An Euler numerical integration scheme with adaptive time steps was employed using $\gamma \equiv 1$, $\alpha = 0.1$ where equilibrium spin structures are determined by averaging the long-time behavior of ${\bf S}_i(t)$.  

  The N$\acute{e}$el order parameter can be calculated as the root mean square of the three sublattice magnetizations as in Ref.\cite{szunyogh09}.  For simplicity, a large single ion anisotropy $D=-1.0$ was assigned (although larger than the experimentally determined value \cite{vajk05}, this difference is not relevant to the present calculation which simply demonstrates the four zero-field Mn spin states) and small interlayer exchange $|J'|=0.01$ and small in-plane anisotropy $|E|=0.01$, we find $T_N \simeq 0.43$. 

 The four principal Mn spin phases result from these simulations by changing the signs of $J'$ and $E$.  An example spin configuration determined at $T=0.40$ is shown in Fig. 1 which correspond to the $\Gamma_1$ phase (${\bf S}_A =- {\bf S}_B$, ${\bf S} \perp {\bf a}$).  Note that the effect of thermal fluctuations is to induce deviations from the idealized structure. The other three principal phases were also observed in the numerical results, dependent on the signs of $J'$ and $E$ , as follows: $\Gamma_2$ with $J'>0$, $E < 0$; $\Gamma_3$ with $J'<0$, $E < 0$; $\Gamma_4$ with $J'<0$, $E> 0$.
%Fig.1.
\begin{figure}[ht]
\includegraphics[width=0.4\textwidth]{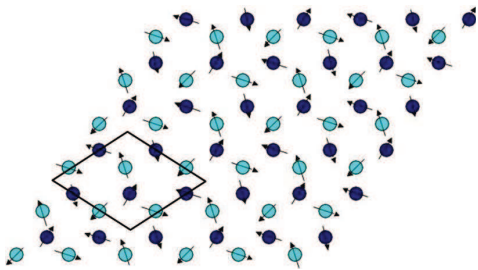}
\caption{ LLG simulation results using $J=1.0$, $J'=0.01$, $D=-1.0$, $E=0.01$ $T=0.4$.  This corresponds to the case of phase $\Gamma_1$.  An outline of the unit cell is shown.}
\end{figure}

  We note that the near-neighbor exchange interaction between Mn spins in HoMnO$_3$ has been estimated to be  $J=2.44 ~ meV \simeq 28 ~ K$ from spin-wave dispersion curves \cite{vajk05}.  Using $s=2$ associated with Mn$^{3+}$, we find $T_N \simeq 0.43 J s^2  \simeq 112 K$, considerably larger than the observed $76 ~ K$.  The reasons for this discrepancy are not clear but may be due to the assumed value of $J'$.   

The crucial role of the Ho-ion spins in driving Mn-spin reorientation transitions is made clear by considering a simple Landau-type free energy, derived from mean field theory, which accounts for the microscopic interactions identified above.  This approach has the advantage of revealing explicitly the competition between the different microscopic energy  and entropy (${\mathcal S}$) contributions through the relation $F=E-T {\mathcal S}$, where thermal effects are calculated in a manner similar to the Brillioun function \cite{bak80,plumer91}.  Single-ion anisotropy associated with the Ho ions  is assumed to be positive ($D_0 > 0$) giving ${\bf S}_0 \parallel {\bf  c} \parallel {\bf {\hat z}}$.  Keeping only relevant terms to sixth order yields (in units of the Boltzmann constant):  
\begin{eqnarray}
F/k_B &=& A S^2 + A_0 S_0^2 + \frac{1}{2} B S^4 + \frac{1}{2} B_0 S_0^4 +  \frac{1}{3} C S^6   \nonumber \\
&+&  \frac{1}{3} C_0 S_0^6 + \tilde{K}S_0S^3\cos(3\phi) + ES^6\cos(6\phi), 
\end{eqnarray}
where $A=a(T-T_N)$ and $A_0=a_0(T-T_{N0})$ $B=bT$, $B_0=b_0T$, $C=cT$, $C_0=c_0T$.  Here, the Mn$^{3+}$ spins order at the N$\acute{e}$el temperature $T_N=-J({\bf Q})s^2/a$.   For the Ho$^{3+}$ ions, the corresponding parameter is $T_{N0}=(-J_0({\bf Q}_0) + D_0)j^2/a_0$ which involves the total angular moment\cite{munoz01} $j=8$. Expressions for the entropy constants are given in Ref.\cite{plumer91} and yield $a=2, b=0.867, c=0.610, a_0=2.667, b_0=1.43$, and $c_0=1.21$.  The remaining model parameters involving the exchange and anisotropy are estimated to reproduce the approximate sequence of phase transitions observed at zero field.  We note that this approach is expected to be increasing less quantitative as the temperature is lowered where the Landau assumption of a small order parameter $S$ breaks down. 

 For spin structures with a periodicity of 3 relative to near-neighbor spacing on a triangular lattice,  $J({\bf Q}) = -3J_1$ (also see below).  For this analysis, we set $T_N = 76 K$ which yields the mean-field estimate $J_1 = 1.15 meV$, more than a factor of two smaller than the measured value.  This difference is consistent with the expected role of spin fluctuations in quasi-2D frustrated antiferromagnets (a suppression of $T_N$). An estimate of $T_{N0}$ can be made using the measured Curie-Weiss behavior \cite{munoz01}, which yields the estimate $T_C=-17 ~ K$.  This value may be compared with the above expression for $T_{N0}$ using $Q=0$ and $J_0(0) = 6J_{01}$ where $J_{10}$ is the near-neighbor exchange interaction between Ho ions in the triangular plane.  For simplicity, we ignore the anisotropy $D_0$ so that the estimate $J_{10} \simeq 0.1 K$ can be made.  This result is consistent with very weak Ho-Ho spin coupling.  Assuming a period-3 near-neighbor modulation of the Ho moments then gives $T_{N0} \simeq 8 K$, associated with a pure Ho system.

  In the free energy (4), $\phi$ represents the angle between Mn spins and the triangular lattice defined by the Ho ions.  The trigonal $\tilde{K}$-term is minimized by $\phi = n \pi /3$, independent of the sign of $\tilde{K}$, whereas the basal-plane anisotropy $E$-term favors $\phi = n \pi /3$ for $E <0$ and $(2n+1) \pi / 6$ for $E>0$. 

Numerical minimization of $F(S,S_0,\phi)$ reveals a variety of phase transition sequences as the temperature is lowered below the paramagnetic phase ($S=S_0=0$), dependent on the assumed values for $E$ and $\tilde{K}$.  Variations occur mainly through the sign of $E$ and the magnitude of $\tilde{K}$. For example, the using $\tilde{K}=25 ~ K$ and $E=1 ~ K$ (for comparison note that $bT_N \simeq 66 ~ K$) yields the results shown in Fig. 2.  The first ordered state is characterized by $S \ne 0, S_0=0$ and $\phi = (2n+1) \pi /6$, corresponding to $\Gamma_4$ or $\Gamma_1$.  In the narrow range between around $T \simeq 37 ~ K$, we find that $\phi$ decreases rapidly to zero ($\phi = n \pi /3$) with the concomitant development of Ho-spin order.  Analysis of the free energy reveals that the Ho ordering occurs at $T_{Ho} = T_{N0} + \tilde{K}^2/(8Ea_0)$.  The narrow region around 37 K (where $0 < \phi < 30^0$) corresponds to phase  $\Gamma_5$ or $\Gamma_6$. In the range $T \simeq 37 ~ K$ to $T \simeq 12 ~ K$, the order corresponds to $\Gamma_2$ or $\Gamma_3$.  Finally, below $T \simeq 12 ~ K$, $\phi$ becomes non-zero again, corresponding to re-entrance to $\Gamma_5$ or $\Gamma_6$. 
 
The trigonal term is allowed by symmetry only if the magnetic order alternates in sign along the c-axis for either Mn or Ho spins (but not both) and is ferromagnetic for the other species. A conclusion of Ref.\cite{nandi08} is that Ho order is antiferromagnetic between triangular planes so that for phases where $S_0 \ne 0$ in the present analysis, the Mn order must be ferromagnetic along the c-axis.  This suggests that the main sequence of phases in HoMnO$_3$ as the temperature increases is given by $\Gamma_6 \rightarrow \Gamma_3 \rightarrow \Gamma_4$, with a narrow region of stability of the intermediate phase $\Gamma_6$ between $\Gamma_3$ and $\Gamma_4$.  This complex sequence of transitions, involving re-entrance of the state $\Gamma_6$, is consistent with experimental results (see, e.g.,  Ref. \cite{lorenz05}).

Other choices of parameter values yields different phase transition scenarios.  For example, a smaller value of $\tilde{K}=10 ~ K$ stabilizes the sequence  $\Gamma_6 \rightarrow \Gamma_4$, with increasing temperature.  With $E<0$, $\phi = n \pi /3$ is stabilized ($\Gamma_2$ or $\Gamma_3$), which yields simultaneous Mn and R ordering at $T_N$. These scenarios have been reported to occur in ErMnO$_3$, YbMnO$_3$ and TmMnO$_3$. 

\begin{figure}[ht]
%Fig. 2
\includegraphics[width=0.45\textwidth]{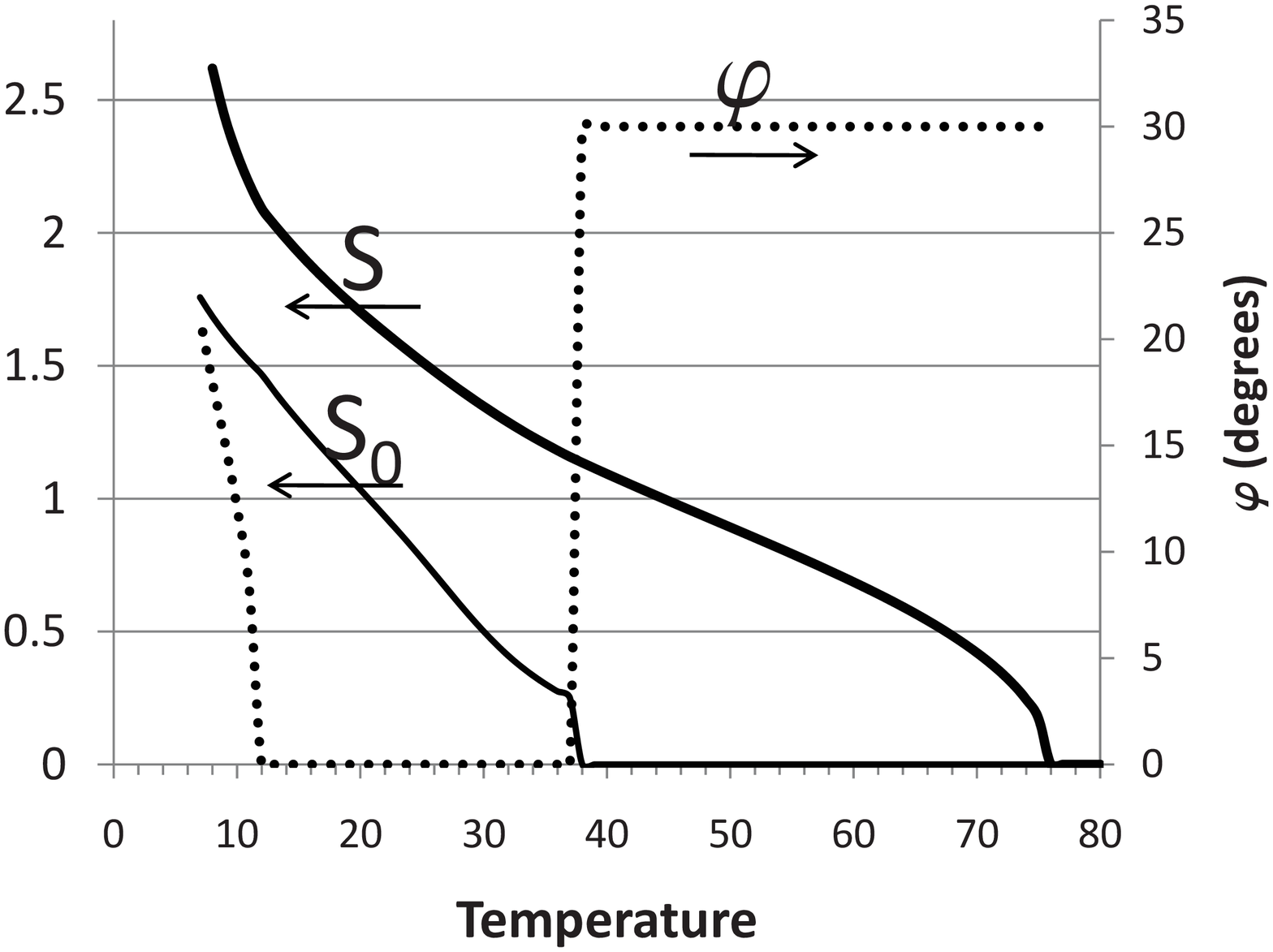}
\caption{ Results of numerical minimization of the mean-field free energy (4).}
\end{figure} 

The pricipal impact of a magnetic field applied along the $c$ axis is the suppression of the $\Gamma_3$ phase ($\phi = n \pi /3$).  A simple analysis illustrates how this can arise within the Landau free energy model. An applied field will induce a uniform contribution to the spin density giving
\begin{eqnarray}
{\bf s}({\bf r}) = {\bf m} + [{\bf S}e^{i{\bf Q} \cdot {\bf r}} +  {\bf S_0}^*  e^{-i{\bf Q_0} \cdot {\bf r}} + c.c.]
\end{eqnarray}
and the free energy functional $F[{\bf s}({\bf r})]$ (with a Zeeman term added) expanded in powers of ${\bf s}$ will contain numerous additional terms involving the magnetization $m = \chi H_z$ \cite{plumer07}.  Some of the additional terms serve to effectively introduce a field dependence to the coefficients which appear in the free energy (4). For example, both anisotropy terms are renormalized as $\tilde{K} \rightarrow K_H = \tilde{K} + \tilde{K}_0 m^2$ and $E \rightarrow E_H = E + E_0m^2$ where $\tilde{K}_0$ and $E_0$ are new parameters.   The stability of the phase $\Gamma_3$ from the numerical minimization of the free energy (4) is found to be very sensitive to the values of both $K_H$ and $E_H$.  For example, with the nominal zero-field value $E_H=1.0$, this phase is stable in the region $12~K \le T \le 37~K$.  A field induced increase to a value $E_H=1.1$ would shrink this to a temperature interval $16~K \le T \le 34~K$ and at $E_H=1.3$ the $\Gamma_3$ phase is suppressed entirely in favor of the $\Gamma_4$ phase.  Similar sensitivity is found with variations in $K_H$. We note that a term of the form $m S^3\cos(3\phi)$ can also occur but only if the Mn spins order anti-ferromagnetically between planes.  This term thus enhances the stability of the $\Gamma_1$ phase.  These results are consistent with experimental magnetic phase diagrams of HoMnO$_3$ where above a relatively small applied field of 3-4 T the $\Gamma_4$ spin structure occupies most of the phase diagram and the $\Gamma_3$ state is absent. In addition, the $\Gamma_1$ phase is observed at low temperatures but only at fields above 1.5 ~ T. At lower temperatures, where the Landau model assumptions are less reliable, we find that the intermediate phase $\Gamma_6$ is stabilized. A wide variety of scenerious have been proposed on the nature of the spin ordering in this region of the phase diagram based on experimental work.

The above analysis was made with the assumption that the spin orderings are commensurate with the lattice.  Due to both small inter-plane coupling of Mn spins as well as the small distortion of the triangular lattice, we argue that the magnetic ordering should in fact be slightly incommensurate. 
We confirmed in the numerical solutions to the LLG equations the expected incommensurabilty of the in-plane modulation driven by interlayer coupling $J'$ of $AB$ stacked triangular layers \cite{saka84}. This was observed as a deviation from $120^0$ in the average Mn inter-spin angle as a function of small $J'$. The effect of the distortion of the triangular lattice (characterized by the deviation of the parameter $x$ from $1/3$) is seen by considering the wave vector which minimizes the Fourier transform of the in-plane Mn exchange interactions, $J({\bf Q})$.  Referring to Ref. \cite{munoz01}, vectors connecting site $0$ to its nearest neighbors are given by ${\bf r}_{10} = (\frac{3}{2}x,x)$, ${\bf r}_{20} = (\frac{3}{2}x,-x)$,  ${\bf r}_{30} = (\frac{3}{2}x - \frac{1}{2},x-1)$, ${\bf r}_{40} = (\frac{3}{2}x - 1,-x)$,       ${\bf r}_{50} = (\frac{3}{2}x - 1,x)$, ${\bf r}_{60} = (\frac{3}{2}x - \frac{1}{2}, 1-x)$  relative to crystallographic vectors ($a{\bf {\hat x}}, b{\bf {\hat y}})$ where $b=\sqrt{3}/2$. In this description, the simple hexagonal lattice is defined by the Ho-ion sites. There are two Mn-Mn bond lengths, $(\sqrt{3x^2})a$ and $(\sqrt{3x^2 - 3x + 1})a$, corresponding to exchange interactions\cite{sato03}, $J_1 > J_2$, giving
\begin{eqnarray}
J({\bf q}) &=& 2J_1 \cos(\frac{3}{2}xq_x)\cos(q_yx) \nonumber \\
&+& 2J_2[\cos(\frac{3}{2}xq_x - \frac{1}{2}q_x) \cos(q_yx - q_y) \nonumber \\
&+& \cos(\frac{3}{2}xq_x - q_x) \cos(q_yx)]
\end{eqnarray}
where $(q_x,q_y)=(aQ_x,bQ_y)$.  With $J_2=J_1$ and $x=\frac{1}{3}$, this function is minimized by near-neighbor spins having a relative angle of $120^0$.   Deviations from this value as a function of either $x$ (with $J_2 = J_1$) or  $J_2/J_1$ (with $x$=0.322, the observed value in HoMnO$_3$ \cite{munoz01}) are shown in Fig. 3. 

\begin{figure}[ht]
%Fig. 3
\includegraphics[width=0.4\textwidth]{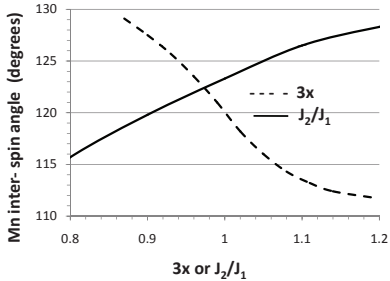}
\caption{ Results of numerical minimization of $J({\bf Q})$ (4) showing the impact of the angle between near-neighbor Mn spins due to the lattice distortion $x$ (with $J_2/J_1=1$) and also due to $J_2/J_1 \ne 1$ (at the value $x$=0.322).}
\end{figure} 

In addition to demonstrating that the four principal Mn-spin states in HoMnO$_3$ are a consequence of a simple spin Hamiltonian, our mean-field analysis has also shown that an unusual trigonal anisotropy term leads to Mn-Ho couplng which accounts for the observed complex phase transition sequence at zero field.  This investigation has revealed explicitly the relation between Ho spin order and Mn spin reorientation.  Extensions of the LLG approach, which is not limited by the same assumptions as Landau theory, to include this coupling and an applied magnetic field to examine phase diagrams and spin excitations are planned.  Our work has also provided strong theoretical arguments which support a scenario where the in-plane modulation characterizing Mn spin order in many RMnO$_3$ compounds should exhibit a slight incommensurability away from the normally assumed period-3 modulation. This deviation from a perfect 120$^0$ spin structure has not yet been reported, likely a consequence of the smallness of the incommensurability, but deserves a more detailed investigation.

We thank  J. Mercer, G. Quirion and S. Curnoe for enlightening discussions. This work was supported by NSERC of Canada and the Atlantic Computational Excellence Network (ACEnet).

\end{document}